# How to Evaluate the Accuracy of Online and AI-Based Symptom Checkers: A Standardized Methodological Framework

Marvin Kopka[1*] & Markus A. Feufel[1]

[1]Division of Ergonomics, Department of Psychology and Ergonomics (IPA), Technische Universität Berlin, Marchstr. 23, 10587 Berlin, Germany

## Abstract

Online and AI-based symptom checkers are applications that assist medical laypeople in diagnosing their symptoms and determining which course of action to take. When evaluating these tools, previous studies primarily used an approach introduced a decade ago that lacked any type of quality control. Numerous studies have criticized this approach, and several empirical studies have sought to improve specific aspects of evaluations. However, even after a decade, a high-quality methodological framework for standardizing the evaluation of symptom checkers remains missing. This article synthesizes empirical studies to outline a framework for standardized evaluations based on representative case selection, an externally and internally valid evaluation design, and metrics that increase cross-study comparability. This approach is backed up by several open-access resources to facilitate implementation. Ultimately, this approach should enhance the quality and comparability of future evaluations of online and AI-based symptom checkers to enable meta-analyses and help stakeholders make more informed decisions.

# Introduction

Symptom checkers (also called 'symptom-assessment applications', 'online symptom checkers', or 'self-assessment applications') are websites or mobile apps in which medical laypeople can enter their symptoms. The apps then provide potential diagnoses and 'self-triage' advice, advising users if and how urgently they should seek care. The first study to systematically analyze the accuracy of these applications was conducted in 2015, and their accuracy has been debated ever since.[1] This seminal study evaluated symptom checkers using 45 medical case vignettes (15 emergency care cases, 15 non-emergency cases and 15 self-care cases) that were taken from various medical resources, including medical education textbooks. The gold standard solution—that is, the most appropriate action for each case—was determined by two physicians who independently rated each case and then discussed disagreements. An unrelated researcher entered all cases into the various symptom checkers, and the authors calculated the proportion of cases correctly solved as the main outcome. This procedure has been used in most subsequent studies, sometimes with slight modifications such as adding more vignettes and triage levels, using lay-friendly phrasing of the vignettes, or including large language models as symptom checkers.[2–5] However, most of these studies acknowledged limitations with this approach and called for improved methods. Systematic reviews that attempted to determine the accuracy of symptom checkers across multiple studies quickly reached the consensus that these methods were often of low quality and that cross-study comparability was limited.[6–9] In recent years, some studies have explicitly formalized this criticism, whereas others have proposed solutions to address it, including several of our own.[6,10–14]

In this article, we do not want to add to this criticism; instead, we present an integrated evaluation framework that can (a) be used to conduct high-quality symptom checker evaluation studies and (b) standardize the evaluation procedure to increase cross-study comparability of symptom checkers and large language models. Because self-triage advice is arguably the most useful information for medical laypeople, we focused this framework to target self-triage accuracy as the main outcome.[15]

# Limitation and Challenges of Previous Methodologies

Most studies evaluating symptom checkers have criticized the existing evaluation approach for being artificial. In particular, the vignettes describe idealized, unambiguous cases, and some include scenarios for which symptom checkers would rarely be consulted. For example, the original vignette set includes a case of recurrent aphthous stomatitis—a condition with recurring mouth ulcers. Although these symptoms may be unexplainable upon first appearance, patients typically become familiar with the condition and can recognize it upon recurrence.[16] As a result, patients would most likely not consult a symptom checker. If the aim is to determine an accuracy metric that can be generalized to real-world interactions and scenarios in which symptom checkers are actually approached, including such cases seems questionable. Apart from vignettes, current evaluation approaches have several other shortcomings that we grouped into four categories: generalizability, symptom input, gold standard assignment, and metrics, see Figure 1. We build on these points to develop our standardized methodological framework.

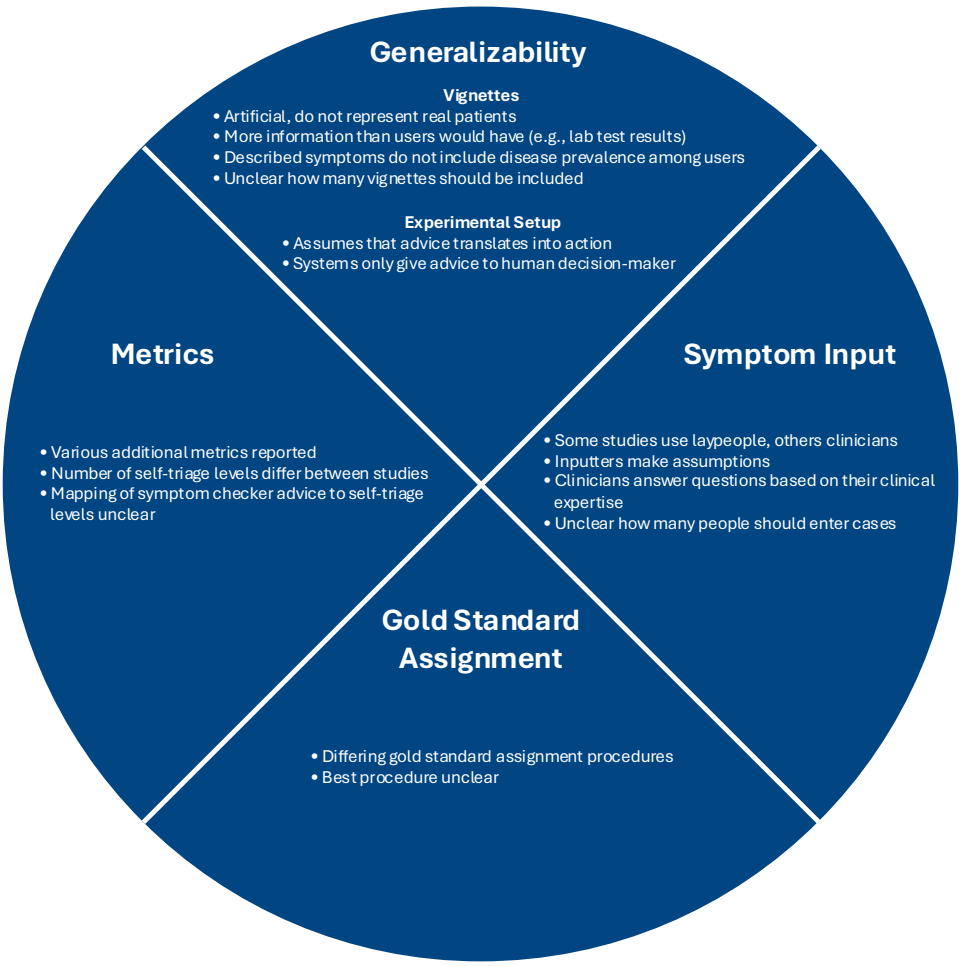

*Figure 1*: Four categories of criticism regarding symptom checker evaluation studies.

The first point concerns the generalizability of the evaluations. This includes both the vignettes and the experimental setup, which, according to ecological validity theory, should resemble real-world use cases and interactions to yield results that can be generalized.[17] Traditional vignettes have been criticized for a lack of representativeness for several reasons. First, they are often derived from medical education textbooks and are therefore artificial, not representing unspecific complaints for which patients would use an SAA.[10,14] Second, these cases are mostly written post hoc by clinicians who have access to more specialized information (e.g., diagnoses, lab test results, clinical examinations) than a patient consulting a symptom checker.[10,13,18] In other words, thus far existing vignettes neither contain nor is it clear which type of information should be included in vignettes to reflect the types of problems actual users of symptom checkers face. Third, the cases described in the vignettes do not reflect the natural base rates of (non-)emergency versus self-care cases among users.[10,13,14] Fourth, there is no consensus on the number of vignettes that should be included in a vignette set or how to ensure their quality.[10] The experimental setup focuses on symptom checker accuracy and has thus been criticized for implicitly assuming that symptom checker advice directly translates into user actions, even though symptom checkers merely give advice to a user that may or may not follow this advice.[19–21] This limitation confines research to assessing only a 'technical accuracy' of a symptom checker, without addressing its likely real-world impact.

The second point concerns the procedure for inputting symptoms. Typically, a single person—who may or may not have medical expertise—enters the symptoms. Because not all information that a symptom checker might ask for is included in the vignette, the inputter must make assumptions about the case when asked about it. Thus, clinicians tend to rely on their clinical judgement and expertise, whereas laypeople—the actual users of symptom checkers—use various strategies, ranging from guessing to ignoring the questions they are asked.[10,11,22] It is also unclear how many inputters should be involved in the evaluation to yield valid performance estimates.[10] These issues suggest that the final output is highly dependent on the inputter(s), which creates an information bias that limits the internal validity of evaluation studies.

The third point relates to the gold standard assignment used to assign the solution to a case vignette. Different studies employ varying procedures: some use a single physician, others use multiple physicians, recordings from clinical encounters (such as telephone triage), or sometimes the authors even determine the gold standard solution themselves.[10,12] This variation not only limits the comparability between studies but also raises questions about the accuracy of the assigned gold standard in some cases.

The last point concerns the metrics used to evaluate symptom checkers. Although most studies report accuracy as the proportion of cases solved correctly, some also include additional metrics such as the tendency to over- or under-triage and the safety of the advice.[14] The exact self-triage classifications differ between studies as well: for example, Semigran et al. used a three-tiered approach including 'emergencies', 'non-emergencies', and 'self-care cases', whereas Hill et al. extended this classification to include '1-day urgent cases'.[1,2] Furthermore, because different symptom checkers use varying classifications as well, it is unclear how their advice should be mapped to the study's triage categories (e.g., whether an urgent care clinic is considered emergency or non-emergency care). These issues ultimately limit cross-study comparability.

## Framework

To address these points, we developed an evaluation framework by integrating available empirical studies on methodological improvements. This framework can be found in Figure 2. It can be used for pre-clinical evaluations to identify symptom checkers that are likely to perform well in clinical trials and real-world evaluations. Once identified, such a tool should nonetheless undergo testing in a three-phase clinical trial similar to pharmaceutical trials.[23] Hence, our framework not only standardizes vignette-based symptom checker evaluations but also makes subsequent clinical trials more cost-efficient by identifying tools likely to yield positive outcomes.

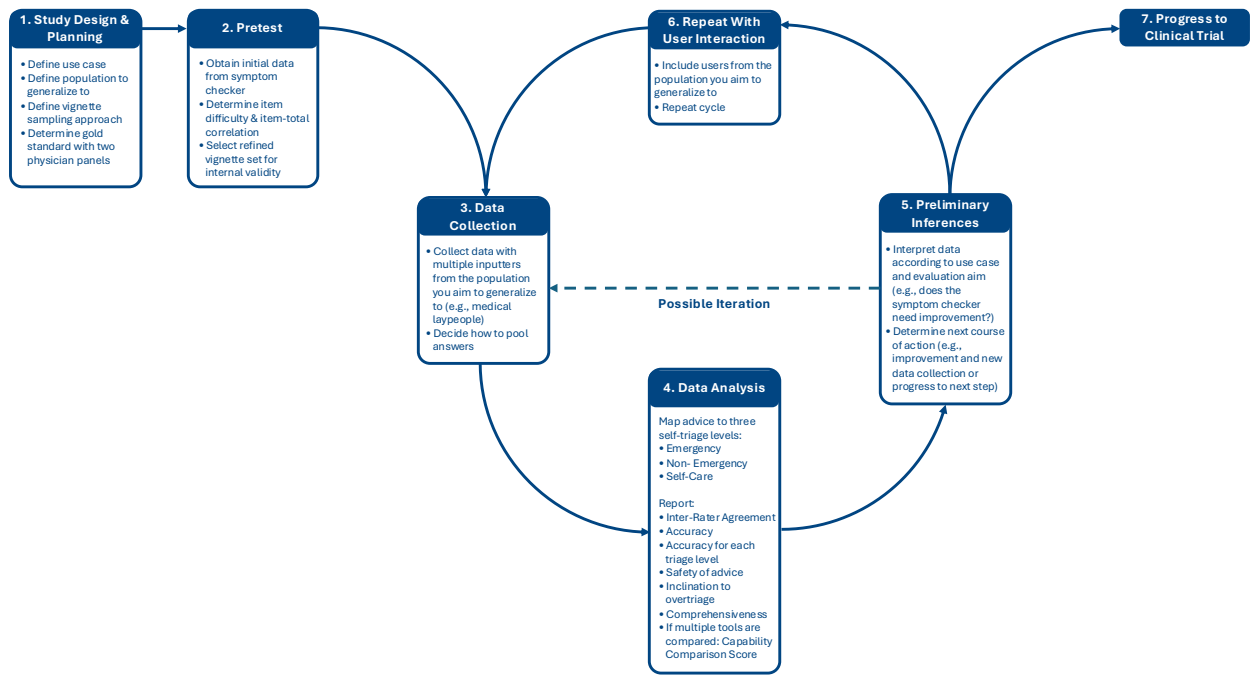

*Figure 2*: Integrated pre-clinical evaluation framework for evaluating the self-triage accuracy of symptom checkers and AI tools.

In the beginning (part one), evaluators should clearly define the use case they intend to examine, such as 'self-triage decisions' or 'emergency care decisions'. Next, they should specify the target population they wish to generalize to. For the self-triage use case, this might include symptom checker users deciding on their next course of action. Then, they should define a vignette sampling approach, which ensures that vignettes are representative of real patient cases and accurately reflect the disease/symptom prevalence relevant to the use case. For example, the approach could sample real patient cases stratified according to the prevalence of symptom types entered into symptom checkers. A systematic sampling procedure to do that is available in the *RepVig Framework,* and for the self-triage use case, a representative vignette set is provided in the framework's validation study.[13] At this stage, researchers should also assign a gold standard solution to each case. According to a study by El-Osta et al., this should involve two physician panels that independently rate the cases in focus groups and resolve any disagreement through discussion until consensus is reached.[12]

Next, evaluators should obtain initial data from some symptom checkers to refine the vignette set according to test-theoretical criteria (part two) to ensure that vignettes are not only externally but also internally valid. This process involves determining the item difficulty for each vignette and excluding cases with an item difficulty of zero (to ensure that vignettes add meaningful information and are not impossible to solve), as well as calculating the item-total correlation and excluding any cases with a negative or zero item-total correlation (to ensure that only cases accurately predicting overall performance are included). A procedure for this is outlined in one of our previous studies.[24] The size of the final vignette set should ultimately be determined using a power analysis. However, given that entering a large number of vignettes may be infeasible, a target sample size of at least 45 vignettes is recommended, as this number has proven to be feasible to produce and can be entered by a single individual within a reasonable timeframe.

Using the refined vignette set, data of all symptom checkers can be collected (part three). Multiple inputters (at least two, possibly more) should enter every case into each symptom checker and select a 'not sure' option in cases of missing information. Their results should then be pooled. This can be achieved using several algorithms, but the best approach appears to be a majority vote, that is, the advice most frequently given to all evaluators.[11] For example, if two inputters receive the advice to seek emergency care while one inputter receives self-care advice, the recommendation should be coded as 'emergency'.

In the next step, the data analysis (part four), evaluators should map the received advice to a three-tiered classification system—'emergency', 'non-emergency', and 'self-care'—to ensure cross-study comparability. We suggest that a '1-day-urgent' category be classified as 'non-emergency', although sensitivity analyses may improve the quality. For example, researchers could first analyze the data using the proposed mapping, and then re-analyze the data using a different mapping (e.g., '1-day-urgent' classified as 'emergency' or as its own category) to assess the stability of the results. After mapping each recommendation, evaluators should first report the inter-rater reliability among all inputters to identify the influence of different inputters, followed by a set of metrics: overall accuracy, accuracy for each triage level, safety of advice, inclination to overtriage, and comprehensiveness.[14,25] These metrics were identified through systematic review of previously reported metrics and can increase comparability across different studies.[14,25] If multiple symptom checkers are evaluated simultaneously, we propose additionally reporting the Capability Comparison Score (developed in a previous study) to determine how symptom checkers perform relative to each other.[14,25] To assist researchers in reporting and visualizing these metrics, the R package symptomcheckR is available.[25]

In the next step, the results should be interpreted according to the defined use case, and the next course of action should be determined (part five): if developers aim to validate their tools, they may either decide to improve their tool and test it again (by going back to step three) using them same setting or continue with the evaluation and test the best performing tools with users in the loop making self-triage decisions (step six). In this phase, users should be provided with the symptom checker, and the tool's impact instead of its 'technical accuracy' should be assessed in a new evaluation with sufficient statistical power.[21] If these results are also promising, the symptom checker can then be tested in a clinical trial with real patients and their symptoms (step seven).

## Open Questions

Our approach leaves several open questions for future research. First, none of the vignettes (neither Semigran's vignettes nor our own) include additional information for questions that symptom checkers may ask. Future research could develop a method, to supplement this missing information—perhaps using a hybrid approach that combines interviews with patients from whom the case vignettes were derived and synthetical AI-generated supplementary data based on these interviews. Second, it remains unclear whether 'accuracy' or a 'correct' solution should be the main outcome. Perhaps a binary classification of correct versus incorrect in such a task that is associated with high uncertainty may be less relevant than assessing the impact of the advice—specifically, whether it is safe and appropriate for the individual and whether it increases or decreases healthcare demands. Third, with the introduction of Large Language Models as an alternative to traditional symptom checkers, output variability plays an even greater role. Future research should address how to manage the variability of generated outputs when provided with identical inputs. Fourth, current evaluations do not specifically include atypical presentations. It remains unclear whether case vignettes are only suitable for typical cases or if vignette sets for atypical cases could also be developed. Although the *RepVig framework* could be used for developing such a vignette set again, assigning a reliable gold-standard solution to atypical cases will be challenging.[13] Lastly, our approach is highly tailored to a self-triage use case. Although it standardizes most aspects of an evaluation, diagnostic use cases may require additional details, such as clinical plausibility, that are not covered by our approach.

## Outlook

The evaluation framework presented in this article addresses all previously raised points of criticism and aims to improve the quality and comparability of future symptom checker evaluations. However, we acknowledge that the presented approach is more resource-intensive than the traditional approach introduced by Semigran et al. and may not be feasible for every evaluation.[1] To aid researchers in integrating these methods in practice, several open resources are available for the presented use case: for example, representative vignettes are openly accessible and free to use[13], a refined vignette set that satisfies test-theoretical criteria is available as well[24], and all metrics can be easily calculated using the open-source symptomcheckR package.[25] We encourage researchers to build on these resources to improve the quality of future evaluations and enhance cross-study comparability.

## Conclusion

In this article, we summarized limitations and challenges of previous studies evaluating symptom checkers using vignettes. In recent years, several empirical studies have addressed most of these limitations individually, yet a unified methodological framework integrating these findings was missing. We have presented a pre-clinical framework upon which future vignette-based symptom checker evaluations can build to address generalizability, input variability, gold standard assignment, and metrics, and point to several open access resources that evaluators can use. By adopting this approach, researchers can identify well-performing tools for more cost-efficient clinical trials, and they can significantly increase the quality and comparability of vignette-based symptom checker evaluation studies and thus enable reliable evidence syntheses. This can help in getting closer to assessing and improving the effectiveness of symptom checkers.

## Declaration of Interests

The authors declare no conflicting interests.